\documentclass[aps,prx,onecolumn,superscriptaddress,groupedaddress]{revtex4}

\usepackage{amsmath}
\usepackage{amsfonts}
\usepackage{amssymb}
\usepackage{bm}
\usepackage{color}
\usepackage{graphicx}
\usepackage{soul}
\usepackage{hyperref}
\usepackage{enumerate}
\usepackage{cancel}
\hypersetup{
    colorlinks=true,
    linkcolor=blue,
    filecolor=magenta,
    urlcolor=cyan,
}

\newcommand{\eq}{\text{eq}}

\begin{document}
\title{Inducing and optimizing Markovian Mpemba effect with stochastic reset}
\author{Daniel M. Busiello}
\thanks{These authors equally contributed to this work.}
\affiliation{Ecole Polytechnique F\'ed\'erale de Lausanne (EPFL), Institute of Physics Laboratory of Statistical Biophysics, 1015 Lausanne, Switzerland}
\author{Deepak Gupta}
\thanks{These authors equally contributed to this work.}
\affiliation{Dipartimento di Fisica `G. Galilei', INFN, Universit\'a di Padova, Via Marzolo 8, 35131 Padova, Italy}
\affiliation{Department of Physics, Simon Fraser University, Burnaby, British Columbia V5A 1S6, Canada}
\author{Amos Maritan}
\affiliation{Dipartimento di Fisica `G. Galilei', INFN, Universit\'a di Padova, Via Marzolo 8, 35131 Padova, Italy}

\begin{abstract}
A \textit{hot} Markovian system can cool down faster than a \textit{colder} one: this is known as the {\it Mpemba effect}. Here, we show that a non-equilibrium driving via stochastic reset can induce this phenomenon, when absent. Moreover, we derive an optimal driving protocol simultaneously optimizing the appearance time of the Mpemba effect, and the total energy dissipation into the environment, revealing the existence of a Pareto front. Building upon previous experimental results, our findings open up the avenue of possible experimental realizations of optimal cooling protocols in Markovian systems.
\end{abstract}

\maketitle

\section{Introduction}

Any {\color{black}strongly ergodic} system coupled to a thermal bath follows its fate and relaxes towards its equilibrium distribution. Relaxation phenomena have been abundantly studied in the field of condensed matter \cite{dattagupta2012relaxation}, originating a plethora of results preluded by the seminal works of Onsager \cite{onsager1931reciprocal,onsager1931reciprocal-2} and Kubo \cite{kubo1966fluctuation}. At present, theoretical findings cover different aspects, ranging from the investigation of single stochastic trajectories \cite{wang2006dynamic,jarzynski2017stochastic} to the role of dynamic \cite{baiesi2013update,maes2017frenetic} and thermodynamic features \cite{hatano2001steady,busiello2019entropy}.

However, most of the works in the literature inspected close-to-equilibrium conditions, being only recently generalized to systems approaching a non-equilibrium stationary state \cite{baiesi2009fluctuations,polettini2013nonconvexity,altaner2016fluctuation}. Some results also underwent subsequent experimental validations, for example using colloidal systems \cite{blickle2006thermodynamics,hoang2018experimental} and molecular machines \cite{hayashi2010fluctuation}.

Nevertheless, the role of initial conditions in relaxation processes remains elusive. Recent works in this direction provided an encompassing framework to explain the emergence of the Mpemba effect \cite{mpemba1969cool,auerbach1995supercooling,jeng2006mpemba} in Markovian systems \cite{ME-1,IME,PRX-ME,chetrite2021metastable} and granular matter \cite{lasanta2017hotter,prasad-1,biswas2021mpemba}. This effect, documented for the first time by Aristotle \cite{ross1924aristotle}, foresees that, under certain conditions, a \textit{hot} system can cool down faster than a \textit{colder} one. Interestingly, last year has witnessed an outstanding experimental verification of this effect using a colloidal particle setup \cite{kumar2020exponentially,natrevphys}. Another study on the role of initial conditions reported an unforeseen {\it asymmetry} in displacement from equilibrium, resulting in {\it faster uphill relaxation} \cite{lapolla2020faster,manikandan2021faster,van2021lower}. This latter phenomenon seems to happen in Langevin systems near stable minima, but not in a general discrete-state Markovian system, as reported in Ref.~\cite{van2021toward}.

At any rate, these recent studies focused solely on how initial conditions influences the relaxation towards equilibrium. According to an equilibration dynamics, the system explores only a limited portion of accessible space, thus mitigating by far the spectrum of possibilities to go from one equilibrium state to another. Nonetheless, actively driving the system out-of-equilibrium opens up the opportunity to explore a much larger portion of space, and consequently allowing for the search of an optimization algorithm to speed up the relaxation process. {\color{black}In fact, the idea of {\it shortcuts to equilibration} has been previously explored in the quantum domain, where generating coherence speeds up the relaxation process \cite{QMS}, and for Brownian particles, through the optimal protocol known as {\it Engineered Swift Equilibration} (ESE) \cite{ESE-1,ESE-2,ESE-3}. Here, we study the efficiency of a specific class of optimal protocols in speeding up the onset of the Mpemba effect in discrete-state systems}.

Stochastic reset is one of the most common driving mechanisms for Markovian systems, whether they are colloidal particles  \cite{evans2011diffusion,gupta2019stochastic} or bio-chemical networks \cite{roldan2016stochastic}. It encompasses an external stochastic forcing which instantaneously resets the system to a desired state with a given rate; after that, the system immediately restart its underlying dynamics. Stochastic reset has been employed to model different processes, such as search strategies \cite{bhat2016stochastic,bressloff2020directed,bressloff2020search}, catastrophic events \cite{plata2020asymmetric}, and antiviral therapies \cite{ramoso2020stochastic} (see more details in a recent review \cite{evans2020stochastic}). Moreover, resetting mechanism displays interesting thermodynamic features in the context of speed limit \cite{gupta2020tighter}, thermodynamic uncertainty relations \cite{pal2021thermodynamic}, work fluctuations \cite{gupta2020work} and entropy production \cite{busiello2020entropy,fuchs2016stochastic}. Lately, stochastic reset has also been implemented experimentally in two independent works \cite{tal2020experimental,besga2020optimal}, hinting at the promise of fruitful realisable applications, especially on relaxation phenomena and cooling procedures.

In this {\color{black}work}, we theoretically elucidate this latter research path, highlighting that resetting a system to a given state we can induce the Mpemba effect in Markovian systems, or eventually optimize it when already present. Specifically, we derive an optimal resetting protocol to reduce the time to cool down a \textit{hot} system at its equilibrium distribution, ending up experiencing a faster cooling than any \textit{colder} counterpart. Further, since applying an external driving on a system has a cost in terms of dissipated energy \cite{busiello2020entropy,busiello2018similarities}, we present a general way to simultaneously minimize the cooling time and the entropy production into the environment due to stochastic reset. Our framework reveals the existence of a {\it Pareto front} in the space of these above competing features, and it applies to any discrete-state Markovian system.

{\color{black}The rest of the paper is organised as follows. In Section~\ref{model}, we discuss the model and compute the probability distribution of the system's state at a given time. Section~\ref{IME-sec} describes the method to induce the Mpemba effect and its optimization over the onset time as well as dissipation is discussed in Section~\ref{OPE-sec}. Finally we conclude in Section~\ref{concl}. Some detailed calculations are relegated to the appendices.}

\section{Model and framework}
\label{model}
We consider a Markovian system characterized by $N$ discrete states. Let the probability to be in a given state $i$ at a time $t$ be $p_i(t)$, for $1\leq i\leq N$. The transition rate from a state $i$ to $j$, when $i\neq j$, is $w_{ji}$, the $(ij)$-th element of the transition rate matrix $\hat{W}$. The diagonal elements are $w_{ii} = - \sum_{j\neq i} w_{ji}$, to ensure probability conservation. The probability $p_i(t)$ evolves according to the master equation, $\partial_t \vec{p}(t)= \hat{W} \vec{p}(t)$, where $\partial_t$ is a partial derivative with respect to time. This dynamics always admits a unique stationary state, provided the transition rate matrix $\hat W$ is irreducible.

When the system is coupled to a thermal bath at temperature $T_{\rm b}$, it eventually relaxes to the equilibrium Boltzmann's distribution, $p^{\rm eq}_i(T_{\rm b}) \propto e^{-\beta_{\rm b} E_i}$, with $\beta_{\rm b} \equiv (k_{\rm B} T_{\rm b})^{-1}$, i.e., the probability to be in the energy state $E_i$. {\color{black}We also assume that the transition rate matrix governing the system's dynamics, $\hat{W}^{\rm eq}(T_{\rm b})$, satisfies detailed balance, hence its} non-diagonal elements are the transition rates $w^{\rm eq}_{ij} = \mathcal{R} e^{-\beta_{\rm b} \left( B_{ij} - E_j \right)}$, where $\mathcal{R}$ is a constant. Here, $B_{ij} = B_{ji}$ is the symmetric energy barrier between the states $i$ and $j$.

The addition of a stochastic reset, with a constant rate $r$, for $t \in [0, t_{r}]$ drives the system in non-equilibrium conditions, while for $t>t_{r}$ the system follows a reset-free dynamics. Hence, its dynamics reads:
\begin{equation}
\partial_t \vec{p}(t) = \left( \hat{W}^{\rm eq} - r \;\Theta(t_{r} - t) \hat{I} \right) \vec{p}(t) + r \;\Theta(t_{r} - t) \vec{\Delta},
\label{dyn}
\end{equation}
where $\Theta$ is the Heaviside step function, $\hat{I}$ the identity matrix, and $\vec{\Delta}$ a vector with all zeros except at the resetting locations, such that $\sum_i \Delta_i = 1${\color{black}, with $\Delta_i \geq 0$, for all $i$}.


\subsection{Markovian Mpemba effect with and without resetting}
\label{IME-sec} 
The standard Markovian Mpemba effect \cite{ME-1} entails the following protocol. Two identical systems are initialized at equilibrium with two thermal baths at different temperatures $T_{\rm H}$ and $T_{\rm C} < T_{\rm H}$, where the subscripts H and C, respectively, indicate \textit{hot} and \textit{cold}. Then, the temperature of both baths is instantaneously quenched to $T_{\rm b} (<T_{\rm C})$. Both systems eventually relax towards $\vec{p}^{\rm eq}(T_{\rm b})$, and the distance from equilibrium can be quantified using many popular measures, namely the Kullback-Leibler divergence \cite{roldan2012entropy}, the $L^1$-norm \cite{kumar2020exponentially}, the $L^2$-norm \cite{shiraishi2018speed}, and the entropic distance \cite{ME-1}. It has been shown \cite{ME-1} that, under certain conditions {\color{black}(see Appendix~\ref{MME})}, the system initialized at $T_{\rm H}$ has a higher relaxation speed, ending up being closer to equilibrium than the one initialized at a colder temperature $T_{\rm C}$, for all times larger than the crossing time $\tau_{\rm c}$. When the relaxation from $T_{\rm H}$ is exponentially faster than the one from $T_{\rm C}$, the phenomenon is called the \textit{strong} Mpemba effect \cite{PRX-ME}. Here, we are interested in estimating how the crossing time, howsoever measured, changes in the presence of stochastic reset. We name it $\tau_{\rm c}^{(r)}$.

If only the \textit{hot} system is driven away from equilibrium through a stochastic reset, according to the dynamics in Eq.~\eqref{dyn}, this $\tau_{\rm c}^{(r)}$ will depend on resetting rate, resetting location, and resetting time-interval, $t_{r}$. To elucidate this dependence, retracing the steps in \cite{ME-1}, it is instructive to solve the full dynamics using a spectral decomposition of the solution. Being $\lambda_k$ the eigenvalues of the equilibrium rate matrix $\hat{W}^{\rm eq}$, and $\vec{v}^{(k)}$ their corresponding right eigenvectors, the solution of Eq.~\eqref{dyn} reads {\color{black}(see detailed derivation in Appendices~\ref{renewal-eqn} and \ref{up-sol-sec}}):
\begin{equation}
\label{soleig}
\vec{p}(t) = \bigg\{\begin{array}{l l}
\vec{p}^{\rm eq}(T_{\rm b}) + \sum_{k\geq 2} a_k^{(r)}(t)~\vec{v}_k~e^{\lambda_k t} & \;\;\; t \leq t_{r}, \\
\vec{p}^{\rm eq}(T_{\rm b}) + \sum_{k\geq 2}{\color{black} a_k^{(r)}(t_{r})~\vec{v}_k~e^{\lambda_k t}} & \;\;\; t > t_{r},
\end{array}
\end{equation}
where we introduce the modified coefficients:
\begin{eqnarray} 
a_k^{(r)}(t) 
 &=&{\color{black}\bigg[\frac{r d_k}{r - \lambda_k} + \bigg(a_k^{\rm (H)} -\frac{r d_k}{r - \lambda_k}\bigg) e^{(\lambda_k-r) t}\bigg] e^{-\lambda_k t}.}
\label{coeff}
\end{eqnarray}
{\color{black}Notice that the second line of Eq.~\eqref{soleig} is the solution of Eq.~\eqref{dyn} for $t>t_r$, and written in the linear combination of eigenvectors of $\hat{W}^{\rm eq}$ and by imposing the matching condition for solutions at time $t=t_r$.}
All eigenvalues are negative, except for the one (which is equal to zero) whose corresponding right eigenvector coincides with the stationary solution. In Eq.~\eqref{soleig}, we ordered them such that $0 = \lambda_1 > \lambda_2 > \dots > \lambda_N$. In Eq.~\eqref{coeff}, coefficients $d_k$ and $a_k^{\rm (H)}$, respectively, depend on resetting locations and initial conditions. Indeed, they satisfy
\begin{eqnarray}
p^{\rm eq}_i(T_{\rm b}) + \sum_{k \geq 2} d_k v^{(k)}_i &=& \Delta_i, \label{delta}\\
p^{\rm eq}_i(T_{\rm b}) + \sum_{k \geq 2} a_k^{\rm (H)}  v^{(k)}_i &=& p^{\rm eq}_i(T_{\rm H}),\label{eqn-eq}
\end{eqnarray}
considering that the system has been initialized in equilibrium with a thermal bath at temperature $T_{\rm H}$ for Eq.~\eqref{soleig}. Conversely, if the system is initialized at $\vec{p}^{\rm eq}(T_{\rm C})$, the corresponding coefficients will be $a_k^{\rm (C)}$ \footnote{Note that the system initialized at the colder temperature, $T_{\rm C}$, is not driven using the stochastic resetting protocol.}. {\color{black}Note that Eq.\eqref{delta} is the expansion of $\vec\Delta$ in the eigenbasis of $\hat W^{\rm eq}$.}

Although Eqs.~\eqref{dyn}--\eqref{delta} holds for any $\vec{\Delta}$, here we specialize to the case of single-state resetting, i.e., $\Delta_i = \delta_{i,i_r}$, being $i_r$ the resetting state, which is more suitable for experimental realizations. We leave for future investigations the complex case in which the system can reset to a mixture of accessible states.

Let us first briefly discuss the condition for the onset of the Markovian Mpemba effect in the absence of resetting. Since eigenvalues are enumerated in decreasing order, the \textit{hot} system undergoes a faster relaxation than the \textit{colder} one when $\big|a_2^{\rm (H)}|_{r=0}\big| < \big|a_2^{\rm (C)}|_{r=0}\big|$\footnote{$a_2^{\rm (H)}|_{r=0}$ and $a_2^{\rm (C)}|_{r=0}$, respectively, are the solutions of Eq.~\eqref{eqn-eq} for given initial conditions $p^{\rm eq}_i(T_{\rm H})$ and $p^{\rm eq}_i(T_{\rm C})$, and resetting rate $r = 0$.}, hence there exists a crossing time $\tau_{\rm c} >0$ (independently of the measure to quantify the distance from equilibrium {\color{black}(see Appendix~\ref{kl-div-sec})}). Notice that this condition is not always met, indeed we analytically show that any two-state system will never achieve it {\color{black}(see details in Appendix~\ref{twolvsec})}. If $a^{\rm (H)}_2 = 0$, the relaxation of the \textit{hot} system is exponentially faster, since the dominant eigenvalue will be $\lambda_3$ instead of $\lambda_2~(>\lambda_3)$, and the \textit{strong} Mpemba effect will take place. The set of states in the probability space whose projection along $\vec{v}_2$ is zero, i.e., those for which the relaxation is dominated by $\lambda_3$, identifies a $(N-2)$-manifold which we name \textit{strong Mpemba space} (SM space) {\color{black}(see discussion in Appendix~\ref{SM-sec-app})}.

\section{Inducing Mpemba effect}
\label{IME-sec}
Here, we show that stochastic reset can favor the onset of the \textit{strong} Mpemba effect in systems for which the latter phenomenon do not appear according to the equilibration dynamics, i.e., for $\big|a_2^{\rm (H)}|_{r=0}\big| > \big|a_2^{\rm (C)}|_{r=0}\big|$. In fact, we derive the conditions that a \textit{hot} system has to fulfill to be driven towards the SM space using an appropriate stochastic resetting protocol, whatever the \textit{hot} and the \textit{colder} temperatures are. In other words, stochastic reset can let any \textit{hot} system satisfying these requirements relax exponentially faster than any \textit{colder} one. This statement actually lies beyond the Mpemba effect, since stochastic reset can provide a driving, independently of the initial conditions, and in a finite time, towards a state in which the relaxation dynamics is dominated by $\lambda_3$.
\begin{figure}[t]
    \centering
    \includegraphics[width = 0.6\columnwidth]{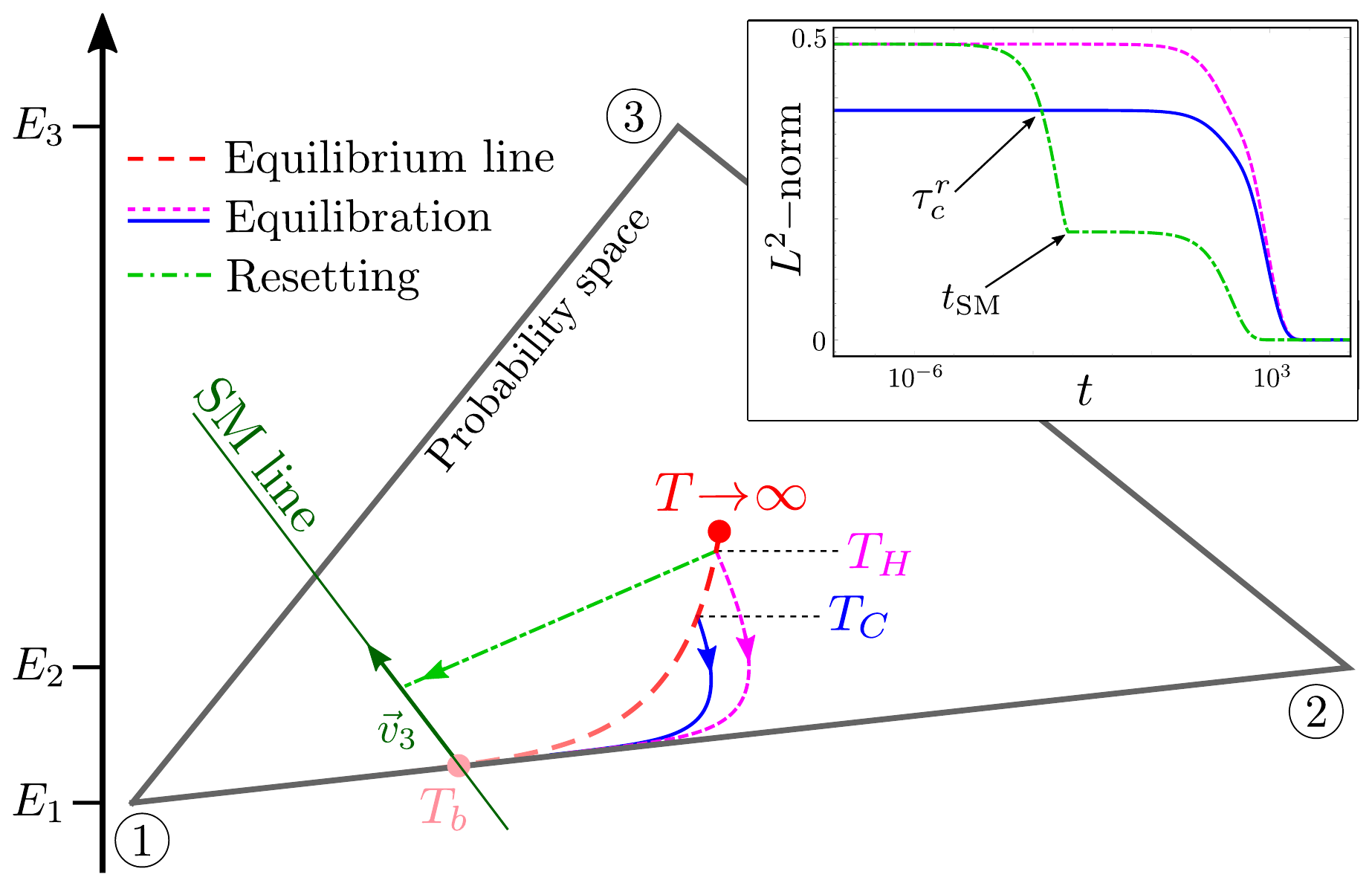}
    \caption{Probability space of a three-state fully connected system obtained by projection on the first two components of $\vec{p}$.
    Equilibration trajectories from $T_{\rm H}$ and $T_{\rm C}$ are the dashed-magenta and solid-blue lines, respectively.
    The trajectory with reset to the optimal state (dot-dashed green line)
    hits the SM space, which is a line for $N = 3$, at time $t_{\rm SM}$. Then, the system relaxes via reset-free dynamics.   
    Here, $i_r^{SM} = 1$, the lowest energy state. The inset shows the $L^2$-norm as a function of time for the same trajectories (with same color coding). The resetting dynamics induces the Mpemba effect at time $\tau_{\rm c}^{(r)} < t_{\rm SM}$. 
    {\color{black}Parameters for this plot are reported in Appendix~\ref{parameters}}.}
    \label{fig1}
\end{figure}
For the purposes of this work, it immediately follows that, if two identical systems properly initialized did not show Mpemba effect, stochastic reset could induce it in its stronger version.

To induce the \textit{strong} Mpemba effect using stochastic reset, we first consider the time to reach the SM space, $t_{\rm SM}$, obtained by imposing that $a_2^{(r)}(t_{\rm SM}) = 0$ (the second coefficient from Eq.~\eqref{coeff} at $t=t_{\rm SM}$), that is {\color{black}(see Appendix~\ref{SM-sec-app})},
\begin{equation}
t_{\rm SM}(r) = \dfrac{1}{r + |\lambda_2|}\ln \bigg[1-\dfrac{(r + |\lambda_2|)a_2^{\rm (H)} }{rd_2}\bigg].
\label{time-sm-2}
\end{equation}
Eq.~\eqref{time-sm-2} admits a physical solution only if $a_2^{\rm (H)}/d_2 \leq 0$. Moreover, the more this ratio is close to zero, the shorter $t_{\rm SM}(r)$ is. For single-state resetting, these two conditions unequivocally determine the optimal state to which stochastically reset the system, $i_{r}^{\rm (SM)}$, in order to reach the SM space as fast as possible. As one could expect, $i_{r}^{\rm (SM)}$ does not depend on $r$. As a first main result, we conclude that it is possible to trigger the onset of the \textit{strong} Mpemba effect through the exploration of the probability space by means of stochastic resetting.

In Fig.~\ref{fig1}, we show a paradigmatic three-state system in which the Mpemba effect is induced using stochastic reset. We also show the dynamical trajectories in the probability space, pinpointing the SM space. We use the $L^2$-norm as a measure of the distance from equilibrium. However, the results do not qualitatively change by employing a different measure {\color{black}(see Appendix~\ref{kl-div-sec})}.

Additionally, we analytically show that the optimal resetting state to have the \textit{strong} Mpemba effect in any two-state system is the lowest energy state {\color{black}(see details in Appendix~\ref{two-state-app})}. Hence, by exploiting a non-equilibrium driving, a discrete two-state system can always experience the Mpemba effect in its stronger version, even if it is forbidden according to the equilibration dynamics. Interestingly, allowing for multi-state resetting will lead to a more complex picture even in simple two-state systems {\color{black}(see details in Appendix~\ref{mix-state})}.

\section{Optimizing Mpemba effect}
\label{OPE-sec}
When a system already exhibits the Markovian Mpemba effect, it can be optimized by reducing the crossing time with respect to the pure equilibration dynamics, and simultaneously minimizing the energy dissipated into the environment due to the non-equilibrium external forcing.

We introduce the following protocol to obtain this two-fold improvement for the relaxation of the \textit{hot} system, independently of the specific values of $T_{\rm H}$, $T_{\rm C}$, and $T_{\rm b}$, considering $T_{\rm H} > T_{\rm C} > T_{\rm b}$:
\begin{itemize}
\item[a)] fix a resetting rate $r$, and a resetting state $i_r$, provided it satisfies the condition $a_2^{\rm (H)}/d_2 < 0$. It allows reaching the SM space in a finite time, $t_{\rm SM}(r)$.

\item[b)] Fix a measure for the distance from equilibrium, and compute the crossing time with resetting, $\tau_{\rm c}^{(r)}$, for a fixed pair $(r,i_{r})$. If $\tau_{\rm c}^{(r)} < t_{\rm SM}$, the crossing happens before the system reaches the SM space. Conversely, if $\tau_{\rm c}^{(r)} > t_{\rm SM}$, the system first overcomes the SM space and then the crossing occurs. The resetting process is stopped at $t_{r} = \tau_{\rm c}^{(r)}$ (see Eq.~\eqref{soleig}), so that for $t>\tau_{\rm c}^{(r)}$ the system dynamics is solely determined by $\hat{W}^{\rm eq}$.
\item[c)] Compute the energy dissipated into the environment up to crossing time $\tau_{\rm c}^{(r)}$, $\Delta S_{\rm env}^{(r)}$ \cite{seifert2012stochastic}:
\begin{eqnarray}
\Delta S_{\rm env}^{(r)} &=& \int_0^{\tau_{\rm c}^{(r)}} \overbrace{\sum_{i<j} w_{ij} p_j(t) \ln \frac{w_{ij}}{w_{ji}}}^{\dot{S}^{(r)}_{\rm env}(t)} \; dt,
\label{senv}
\end{eqnarray}
where $w_{ij}$ are elements of the transition rate matrix with reset governing Eq.~\eqref{dyn}, for $t > 0$. The probability $\vec{p}(t)$ is detailed in the first line of Eq.~\eqref{soleig}.
\item[d)] Build the following functional:
\begin{equation}
F(\gamma,r,i_r) = \gamma \Delta S_{\rm env}^{(r)} + (1-\gamma) \tau_{\rm c}^{(r)}
\label{functional}
\end{equation}
with $\gamma \in [0,1]$. When $\gamma \to 0$, $F \to \tau_{\rm c}^{(r)}$, whereas $F$ weights $\Delta S_{\rm env}^{(r)}$ more than $\tau_c^{(r)}$, as $\gamma$ approaches $1$.

\item[e)] Retrace the steps a)-d) changing $r$ and $i_r$, and find the optimal pair $(r^{\rm opt},i_r^{\rm opt})$, as:
\begin{equation}
    (r^{\rm opt}(\gamma),i_{r}^{\rm opt}(\gamma)) = \arg\min_{r,i_{r}} F(\gamma, r, i_{r}).
    \label{argmin}
\end{equation}
Note that $r$ and $i_{r}$ univocally determine the crossing time $\tau_{\rm c}^{(r)}$, hence there is no need to optimize over it. Clearly, $r^{\rm opt}$ and $i_{r}^{\rm opt}$ are functions of $\gamma$, as they depend on the relative weight of $\Delta S_{\rm env}^{(r)}$ and $\tau_{\rm c}^{(r)}$ while performing the minimization of $F$.
\end{itemize}
Following this protocol, we determine the optimal resetting rate, $r^{\rm opt}$, and state, $i_{r}^{\rm opt}$, so that the system simultaneously minimizes the crossing time and the total dissipated energy with a desired relative weight, $\gamma$. If we were mostly interested in speeding up the Mpemba effect, $\gamma$ would be close to zero. Conversely, $\gamma \to 1$ is associated with minimal dissipation protocols \footnote{As an additional remark, we notice that $i_{r}^{\rm opt}(\gamma)$ does not necessarily coincide with $i_{r}^{\rm (SM)}$.}.

Since the Markovian Mpemba effect can happen even in systems following equilibration processes, at a given crossing time $\tau_{\rm c}$ and with a certain amount of dissipation into the environment $\Delta S_{\rm env}^{\rm eq} = \lim_{r \to 0} \Delta S_{\rm env}^{(r)}$, stochastic reset leads to a two-fold improvement when:
\begin{align}
&\textrm{(i)  } \tau_{\rm c}^{(r)}(r^{\rm opt}(\gamma), i_{r}^{\rm opt}(\gamma)) < \tau_{\rm c} \nonumber \\
&\textrm{(ii)  } \Delta S_{\rm env}^{(r)}(r^{\rm opt}(\gamma), i_{r}^{\rm opt}(\gamma)) < \Delta S_{\rm env}^{\rm eq}. \nonumber
\end{align}

\subsection{Minimizing dissipation and crossing time: a Pareto front}

\begin{figure}[t]
    \centering
    \includegraphics[width = 0.6\columnwidth]{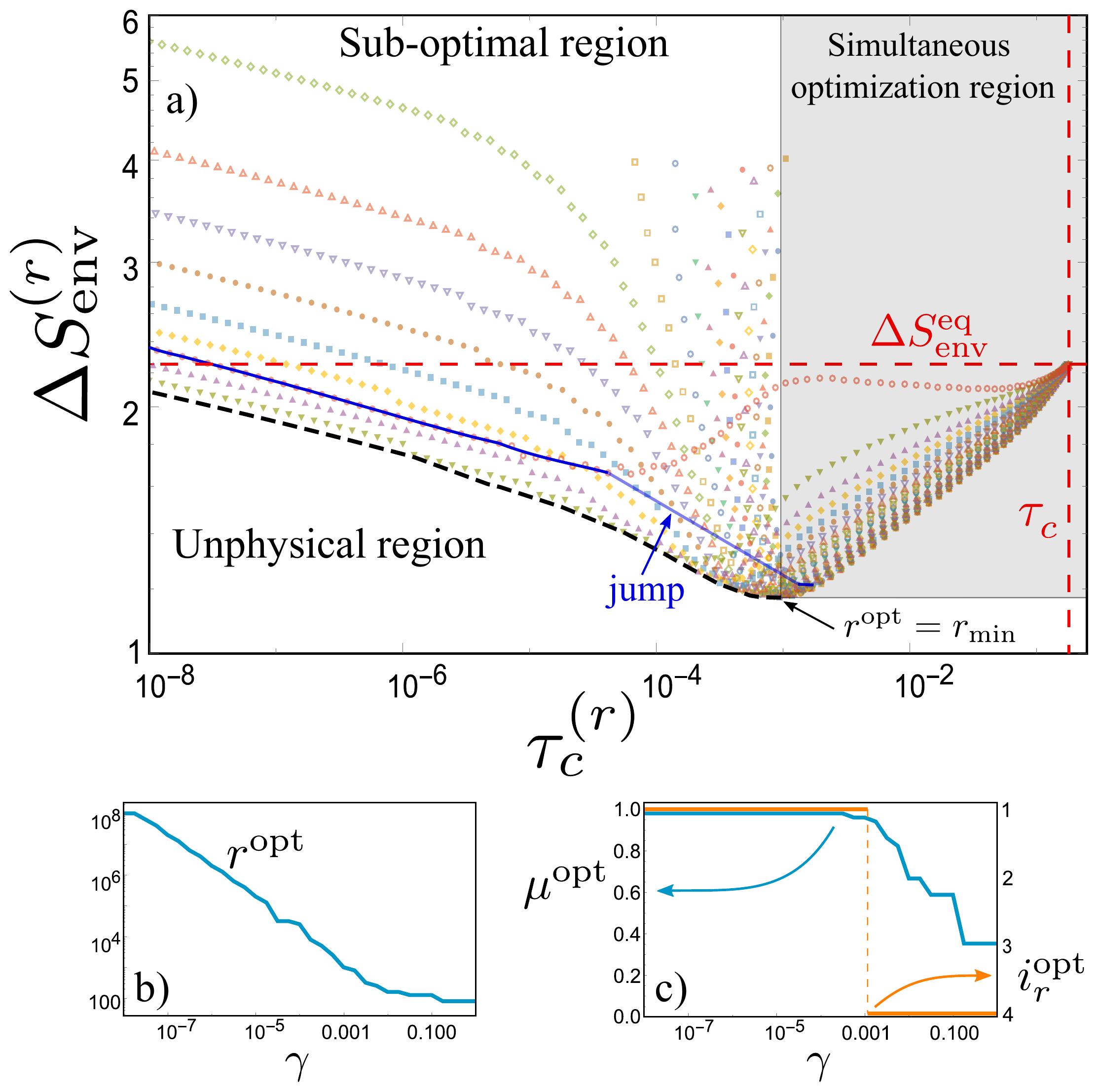}
    \caption{a) Optimal front (blue solid line) in the space $(\tau_{\rm c}^{(r)},\Delta S_{\rm env}^{(r)})$ obtained using single-state resetting, exhibiting a jump. The black dashed line indicates the Pareto front when $\Delta_i = \mu \delta_{i,1} + (1-\mu) \delta_{i,4}$. Each point of the front minimizes $F$ for a fixed $\gamma$ (see Eq.~\eqref{argmin}). Sub-optimal protocols have been obtained for $\mu \neq \mu^{\rm opt}$ and lie above the front (different symbols correspond to different $\mu$'s). Below it, there are unphysical protocols. Stochastic reset can lead to a two-fold optimization, i.e., $\Delta S_{\rm env}^{(r)} < \Delta S_{\rm env}^{\rm eq}$ and $\tau_{\rm c}^{(r)} < \tau_{\rm c}$. For protocols lying in the shaded area, $\tau_c^{(r)}$ and $\Delta S_{\rm env}^{(r)}$ can be simultaneously minimized, resulting in one single optimal point $r^{\rm opt} = r_{\rm min}$. b) Optimal resetting rate as a function of $\gamma$. c) Optimal $\mu^{\rm opt}$ and $i_r^{\rm opt}$ as a function of $\gamma$. {\color{black}For parameters of the plot, see Appendix~\ref{parameters}}. } 
    \label{fig2}
\end{figure}

Evidently, the crossing time and the dissipation rate, $\dot{S}_{\rm env}$, are two competing features. Indeed, a large resetting rate, $r$, brings the system to its resetting location very frequently, leading to a higher rate of dissipation. At once, the crossing time $\tau_{\rm c}^{(r)}$ reduces with increasing $r$. Conversely, when $r \to 0$, the hot system takes a longer time to cross its colder counterpart, since $\tau_{\rm c}^{(r)} \to \tau_{\rm c}$ from below. However, the rate of dissipation due to slow resetting mechanism decreases with decreasing $r$. Here, we aim at minimizing the total dissipation, whose integration domain shrinks as $\tau_{\rm c}^{(r)}$ decreases, suggesting the presence of a minimum for $\Delta S_{\rm env}^{(r)}$ at $r = r_{\rm min}$. This picture prompts the existence of a complex Pareto front by varying the weight $\gamma$ \cite{sole}. Indeed, we have a region in which $\Delta S_{\rm env}^{(r)}$ and $\tau_{\rm c}^{(r)}$ cannot be simultaneously minimized, hence leading to an optimal front, with $r^{\rm opt} > r_{\rm min}$. There also exists a region in which the simultaneous optimization is possible: when $\gamma \to 1$, $r^{\rm opt} \to r_{\rm min}$, and reducing further the value of $r$ both $\Delta S_{\rm env}^{(r)}$ and $\tau_c^{(r)}$ increase. The former leads to a proper Pareto optimality, where the points below the front represent unphysical protocols, while the points above it are sub-optimal protocols. The latter region can be mapped in only one point on the Pareto front.


In Fig.~\ref{fig2}a, we report the optimal front for a four-state fully connected system, showing that it experiences a sudden jump, corresponding to the switch of $i_r^{\rm opt}$ from $1$ to $4$ (see Fig.~\ref{fig2}c). Introducing $\Delta_i = \mu \delta_{i,1} + (1-\mu) \delta_{i,4}$, we resemble a continuous Pareto front, and see that this slight complication with respect to single-state resetting already lead to the emergence of additional features \footnote{Note that in this case, the optimization procedure is decribed by $(r^{\rm opt}(\gamma),\mu^{\rm opt}(\gamma)) = \arg\min_{r,
\mu}F(\gamma, r, \mu).$}. The shaded area represents the region where simultaneous optimization is possible: all points lying in this area will eventually collapse to the single point $r^{\rm opt} = r^{\rm min}$ when optimized. Fig.~\ref{fig2}b-c respectively show the optimal resetting rate and $\mu^{\rm opt}$. We remark that it is possible to reach lower dissipation, and lower crossing time than the equilibrium case, by leveraging the designed protocol (even for the case of single-state resetting). {\color{black}In Appendix~\ref{3-state-PF}, we also illustrate the Pareto front for a three-state system.}

{\color{black}In a nutshell, accelerating the onset of the Mpemba effect is a doable task through stochastic reset, by minimizing only the second term of the functional in Eq.~\eqref{functional}, $\tau_c^{(r)}$, i.e. for $\gamma = 0$. However, as for the ESE \cite{ESE-1,ESE-2,ESE-3}, this speed up comes with an energetic cost, quantified by $\Delta S_{\rm env}^{r}$, i.e. the first term of the functional. Hence, while designing an optimization protocol through the minimization of $F$, $\gamma$ balances the requirements of supporting a small dissipation and obtaining a fast Mpemba effect: the closer $\gamma$ is to $1$, the cheaper the protocol is, in terms of dissipated energy. Remarkably, in glaring contrast with ESE, we show that, at least for some simple systems, the minimum possible achievable dissipation is lower than the one without resetting, $\Delta S_{\rm env}^{\rm eq}$, and it is accompanied by a crossing time (which dictates the onset of Mpemba effect) shorter than $\tau_c$, as resulting from the simultaneous optimization region in Fig.~\ref{fig2}. This observation opens up the perspective of a possible two-fold optimization even in more complex systems.}



\section{Conclusion}
\label{concl}
In this {\color{black}paper}, we found that discrete-state Markovian Mpemba effect can be induced, when absent, or accelerated, when present, using stochastic single-state resetting protocols. Moreover, optimal protocols can be designed to provide a two-fold improvement to this effect, both in the crossing time, and in the inescapable environmental dissipation. Our result evidences the existence of a complex Pareto front in the space of these latter competing features. This highlights the potentialities of non-equilibrium external forcing in empowering the efficacy of spontaneous processes. We also stress that we only dealt with relatively simple protocols which are, in principle, experimentally realizable. More complex resetting mechanisms might hide several intricacies, as discussed here and there in the main text, both theoretically and experimentally, hence deserving a proper investigation.

{\color{black}{\color{black}In particular, herein we discuss two simple settings in which our analytical findings may be experimentally verified and implemented.} i) A Brownian particle can be confined in a double well potential (mimicking a two-state system), {\color{black}and the Mpemba effect has already been observed in this class of systems \cite{kumar2020exponentially}}. At a random interval of time, the particle can be brought to a particular well using an optical trap, e.g. following the techniques introduced in Ref.~\cite{tal2020experimental,besga2020optimal}, {\color{black}hence realizing a stochastic reset at a desired rate}. ii) {\color{black}A similar framework can be designed using single DNA hairpins with both ends attached to optically confined beads, as in Ref~\cite{DNA}. It has been shown} that single DNA hairpins exhibit a double well behavior at a particular end-point separation {\color{black}length}. Therefore, by moving one of the optical traps to a certain end-point separation at random interval of time, one can obtain a single-well behavior, so effectively resetting the system to one single state.}



These results have been presented here for discrete-state systems. Since it has been recently shown that discrete-state and continuous-state systems may exhibit remarkably different equilibration behaviors \cite{van2021toward}, it would be interesting to investigate Langevin systems, as the ones discussed in Refs.~\cite{lapolla2020faster,busiello2019hyperaccurate}, in the presence of stochastic resetting.


\begin{acknowledgements}
D.G. and A.M. acknowledge the support from University of Padova through “Excellence Project 2018” of the Fondazione Cassa di Risparmio di Padova e Rovigo. {\color{black}We thank the anonymous reviewers for their valuable suggestions that have improved the manuscript's content and presentation.}
\end{acknowledgements}

\appendix

\section{Markovian Mpemba effect in the absence of resetting}
\label{MME}
We consider a discrete state (continuous time) Markov system where the state probability of the system evolves according to the following master equation:
\begin{align}
    \partial_t\vec p=\hat{W}^{\rm eq}\vec{p},\label{ME}
\end{align}
where $\hat{W}^{\rm eq}$ is the transition rate matrix with off-diagonal elements:
\begin{align}
    w_{ij}^{\rm eq} \equiv \mathcal{R} e^{-\beta_{\rm b}(B_{ij}-E_j)},\label{rates}
\end{align}
with $\mathcal{R}$ being a constant. The $i$-th diagonal element $w^{\rm eq}_{ii}\equiv -\sum_{k\neq i}w^{\rm eq}_{ki}$ accounts for the exit rate from state $i$, and ensures probability conservation. Here, $w^{\rm eq}_{ij}$ is
the transition rate for the system to hop from state $j$ to state $i$. $B_{ij}=B_{ji}$ is the symmetric barrier between state $i$ and $j$, $E_i$ is the energy of state $i$, and $\beta_{\rm b}\equiv (k_{\rm B} T_{\rm b})^{-1}$ is the inverse temperature of the heat bath coupled to the system. 

The solution of Eq.~\eqref{ME} is
\begin{align}
    \vec p(t) = a_1 \vec v_1 +\sum_{k\geq 2} a_k \vec v_k e^{\lambda_k t},\label{sol-ME}
\end{align}
where $\lambda_k$ is the $k$-th eigenvalue corresponding to $k$-th right eigenvector: $\hat{W}^{\rm eq}\vec v_k = \lambda_k \vec v_k$. Note that left and right $k$-th eigenvectors have the same eigenvalue $\lambda_k$. Moreover, $\lambda_1 = 0$, and $\lambda_k<0$ for $k\geq 2$. The eigenvalue $\lambda_1=0$ stems from the condition that the probability distribution is normalized. Thus, from Eq.~\eqref{sol-ME} we find that the stationary state (ss) corresponds to the zeroth-eigenvalue (since other contributions go to zero in the long-time limit) and that state we call $\vec p^{\rm (ss)}$. Also from the rates \eqref{rates}, we see that the product of the forward transition rates is equal to that of in the reverse direction (which is the detailed balance condition), hence the system eventually reaches a stationary distribution which is the equilibrium distribution $\vec p^{\rm eq} = \vec p^{\rm (ss)}$ from any given initial state. In this case, the equilibrium distribution is the Gibbs-Boltzmann distribution:
\begin{align}
    p^{\rm eq}_i \equiv \dfrac{e^{-\beta_{\rm b} E_i}}{Z(\beta_{\rm b})} \qquad \text{for} \quad i =1,2,3, \dots,
\end{align}
where $Z(\beta_{\rm b})\equiv \sum_j e^{-\beta_{\rm b} E_j}$ is the partition function.

Since in the long run the system will eventually reach the equilibrium distribution $\vec p^{\rm eq}$, the coefficient $a_1 = 1$ and $\vec v_1 =\vec p^{\rm eq}$ in \eqref{sol-ME}. Therefore, the solution can be rewritten as 
\begin{align}
    \vec p(t) = \vec p^{\rm eq} +\sum_{k\geq 2} a_k \vec v_k e^{\lambda_k t},\label{sol-ME-2}
\end{align}
where the coefficients $a_k$ are obtained by imposing the initial condition. Evidently, $\vec p(t\to \infty)\to \vec p^{\rm eq}$.

Now consider that the system starts from a temperature $T_{\rm I}$ (which not necessarily equal to $T_{\rm b}$). Therefore, the initial state is   
\begin{align}
    \vec p^{\rm (I)} \equiv \bigg[\dfrac{e^{-\beta_{\rm I} E_i}}{Z(\beta_{\rm I})},\dfrac{e^{-\beta_{\rm I} E_2}}{Z(\beta_{\rm I})},\dots,\bigg]^\top,\label{IC}
\end{align}
for the inverse temeprature $\beta_{\rm I} \equiv (k_{\rm B} T_{\rm I})^{-1}$. Substituting \eqref{IC} in Eq.~\eqref{sol-ME-2} at time $t=0$, we find the coefficients as function of temperature $T_{\rm I}$: $a_k^{\rm (I)}$. 

Further, the onset of the Markovian Mpemba effect for a system starting from a hot temperature $\beta_{\rm H}$, and reaching the bath temperature $\beta_{\rm b}>\beta_{\rm H}$ faster than the same system initialized at a colder temperature $\beta_{\rm C}<\beta_{\rm b}$ (also $\beta_{\rm H}<\beta_{\rm C}$) is determined by the absolute value of the second coefficient, i.e.,  if \begin{align}
|a_2^{\rm (H)}|<|a_2^{\rm (C)}| \label{cond-1}
\end{align} then the Markovian Mpemba effect occurs.

\section{{\color{black}Solution of Eq.~\eqref{dyn}~for $t\leq t_r$}}
\label{renewal-eqn}
In this section, we consider a system {\color{black}[given in Eq.~\eqref{dyn} for $t\leq t_r$]} which stochastically reset to some particular state $i_r$. For simplicity, we assume that the time to reset is drawn from exponential distribution $r e^{-r t}$, where $r$ is a constant rate of resetting. Thus, the master equation is 
\begin{align}
    \dot p_i=\sum_{j=1}^{N}\hat{W}^{\rm eq}_{ij}p_{j}-rp_i+r\delta_{i,i_r},\label{res-ME}
\end{align}
where the first term on the right hand side is contribution in the absence of resetting, the second (with the minus sign) and third terms, respectively, correspond to loss from the state $i$ and gain to the probability at state $i=i_r$. 

In the matrix form, we rewrite the above equation \eqref{res-ME} as
\begin{align}
    \dot{\vec p}= [\hat{W}^{\rm eq}-r\mathcal{I}] \vec p+r\vec\Delta,\label{res-1}
\end{align}
where $\mathcal{I}$ is the identity matrix, and $\vec\Delta$ is a column vector indicating where the system is resetting, with all zero entries except at the resetting location. 

Let us first consider the case when there is no resetting $r=0$ the above equation becomes {\color{black}$ \dot{\vec p}^{(0)}= \hat{W}^{\rm eq} \vec p^{(0)}$}. The solution of this equation is 
\begin{align}
    \vec p^{(0)}(t)=e^{\hat{W}^{\rm eq}t}{\color{black}\vec p(0)},\label{non-res-1}
\end{align}
{\color{black}where the superscript `0' indicates $r=0$ case}. 

Now let us move our attention for the case when $r\neq 0$, i.e., Eq. \eqref{res-1}, and its solution is 
\begin{align}
    \vec p(t)&
=e^{-r t} e^{\hat{W}^{\rm eq}t} \vec p(0)+r \int_0^t~dt'~e^{-rt'}e^{\hat{W}^{\rm eq}t'}\vec\Delta.
\end{align}

We further assume that the initial conditions for the resetting and non-resetting are identical. Therefore, using Eq.~\eqref{non-res-1}, we rewrite the solution as follows:
\begin{align}
    \vec p(t)=e^{-r t} \vec p^{(0)}(t)+r \int_0^t~dt'~e^{-rt'}e^{\hat{W}^{\rm eq}t'}\vec\Delta. \label{del-eqn}
\end{align}
This is the renewal equation for the resetting distribution. On the right-hand side, the first term accounts for the fact that there is not a single resetting event happened until time $t$ (and the probability for that is $e^{-r t}$) and the second term indicates that the last reset happened with probability $re^{-rt'}dt'$ within the time interval $t'$ and $t'+dt'$.  Eq.~\eqref{del-eqn} is the full solution of the master equation \eqref{res-ME}.

Now, we rewrite the renewal equation \eqref{del-eqn} for the resetting distribution using the eigenvectors and eigenvalues of $\hat{W}^{\rm eq}$ matrix. Since the eigenvectors of $\hat{W}^{\rm eq}$ form a complete basis, we write the vector $\vec \Delta $ as a linear combination of these eigenvectors: $\vec \Delta= \vec p^{\rm eq}+\sum_{n\geq 2}d_n \vec v_n $, and similarly, we have 
$\vec p^{(0)}(t)=\vec p^{\rm eq}+\sum_{n\geq 2}a_n \vec v_n e^{\lambda_n t} $ {\color{black}(see Eq.~\eqref{sol-ME-2})}.
Substituting these in the above equation, we get
\begin{align}
    \vec p(t)&=e^{-r t}\bigg[ \vec p^{\rm eq}+\sum_{n\geq 2}a_n \vec v_n e^{\lambda_n t}\bigg] +r \int_0^t~dt'~e^{-rt'}e^{\hat{W}^{\rm eq}t'} \bigg[\vec p^{\rm eq}+\sum_{n\geq 2}d_n \vec v_n \bigg].
\end{align}
Using the property, $\big(\hat{W}^{\rm eq}\big)^n \vec{p}^{\rm eq}=0$ and $e^{\hat{W}^{\rm eq} t}\vec v_n=e^{\lambda_n t}\vec v_n$, we get
\begin{align}
    \vec p(t)&=e^{-r t}\bigg[ \vec p^{\rm eq}+\sum_{n\geq 2}a_n \vec v_n e^{\lambda_n t}\bigg] +r \int_0^t~dt'~e^{-rt'}\bigg[\vec p^{\rm eq}+\sum_{n\geq 2}d_n \vec v_n e^{\lambda_n t'}  \bigg]\\
     &=\vec p^{\rm eq}+\sum_{n\geq 2}a_n \vec v_n e^{-(r-\lambda_n) t}+\sum_{n\geq 2} \dfrac{r[1-e^{-(r-\lambda_n)t}]}{r-\lambda_n} d_n \vec v_n 
\end{align}

Simplifying the above equation, we get
\begin{align}
     \vec p(t)&=\overbrace{\vec p^{\rm eq}+\sum_{n\geq 2} \dfrac{r d_n }{r-\lambda_n} \vec v_n}^{\equiv p_{\rm ss}^{(r)}}+\sum_{n\geq 2}\bigg[a_n-\dfrac{r d_n}{r-\lambda_n}\bigg] \vec v_n e^{-(r-\lambda_n) t}\label{part-1}\\
     &=\vec p^{\rm eq}+\sum_{n\geq 2} \underbrace{\bigg\{\dfrac{r d_n }{r-\lambda_n}e^{-\lambda_n t} +\bigg(a_n-\dfrac{r d_n}{r-\lambda_n} \bigg)e^{-r t}\bigg\}}_{\equiv a_n^{(r)}(t)}~\vec v_n e^{\lambda_n t},\label{part-2}
\end{align}
where the term $p_{\rm ss}^{(r)}$ in Eq.~\eqref{part-1} is the steady state in the presence of resetting. In the above equation~\eqref{part-2}, $\vec p(t)$ {\color{black}(which is first line of Eq.~\eqref{soleig})} is expressed as a linear combination of eigenvectors of $\hat{W}^{\rm eq}$ with modified coefficients that are time-dependent in contrast to the case for $r=0$. Specifically, the term in the braces in Eq.~\eqref{part-2}
\begin{equation}
    a_n^{(r)}(t) \equiv \left( a_n - \frac{r d_n}{r - \lambda_n} \right) e^{-r t} + \frac{r d_n}{r - \lambda_n} e^{-\lambda_n t} 
    \label{anRt}
\end{equation} 
is the $n$-th coefficient at time $t$ associated with a system with a resetting dynamics at a rate $r$.

{\color{black}\section{Solution of Eq.~\eqref{dyn} for $t>t_r$}
\label{up-sol-sec}
In this section, we write the solution of Eq.~\eqref{dyn} for $t>t_r$, and it can be written in the linear combination of eigenvectors of $\hat{W}^{\rm eq}$, i.e., 
\begin{align}
    \vec p(t) = \vec{p}^{\rm eq} + \sum_{n\geq 2} b_n~\vec{v}_n~e^{\lambda_n (t-t_r)}.\label{up-sol}
\end{align}
Now we aim to find the coefficients $b_n$ by imposing the matching condition, i.e., the solution for $t\leq t_r$ and $t>t_r$ should match at $t=t_r$. This gives (from Eqs.~\eqref{part-2} and \eqref{up-sol}) 
\begin{align}
    b_n = a_n^{(r)}(t_r)~e^{\lambda_n t_r}.\label{bn-eqn}
\end{align}
Therefore, substituting $b_n$ from above Eq.~\eqref{bn-eqn} into Eq.~\eqref{up-sol} gives the second line of Eq.~\eqref{soleig}.
}
\begin{figure}[t]
    \centering
    \includegraphics[width = 6cm]{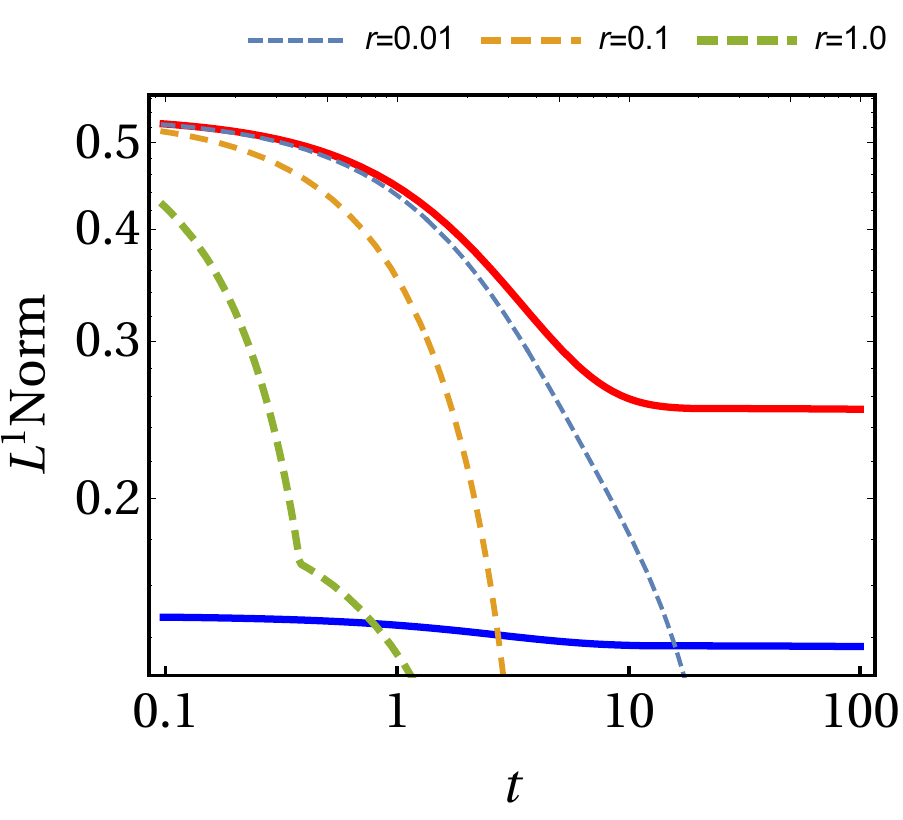}~~
    \includegraphics[width = 6cm]{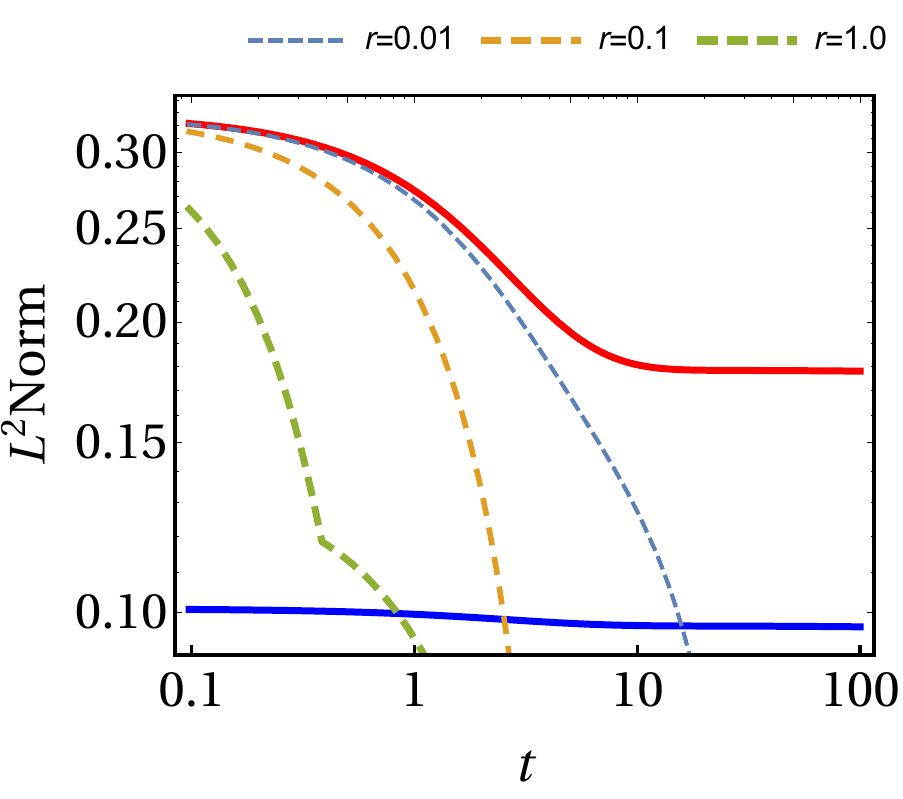}~~
    \includegraphics[width = 6cm]{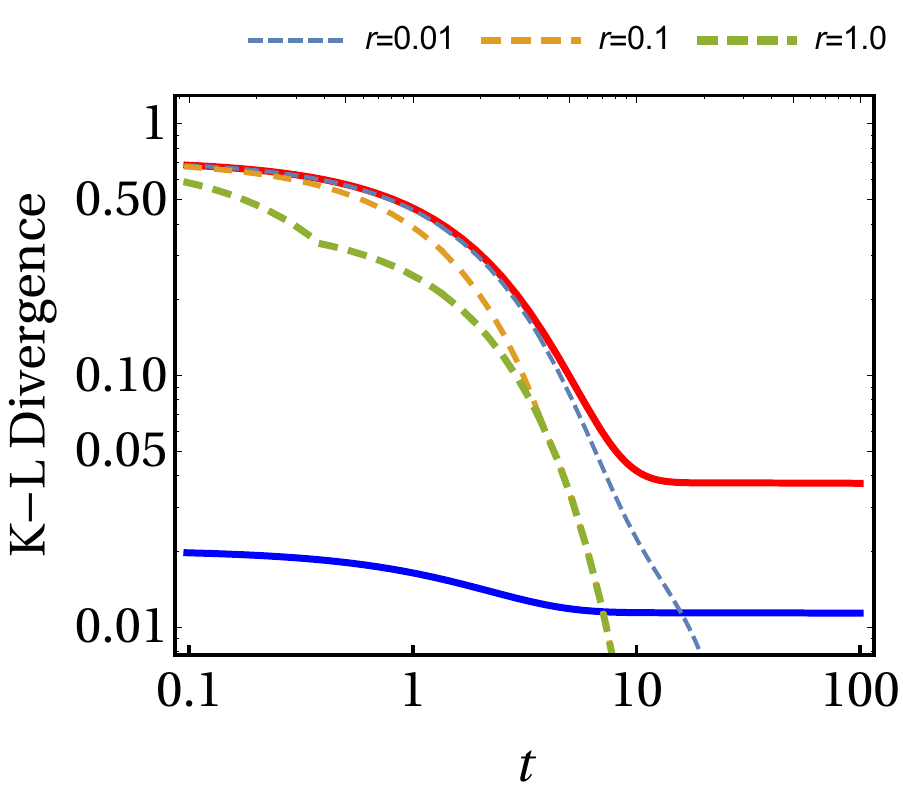}\\
    \includegraphics[width = 6cm]{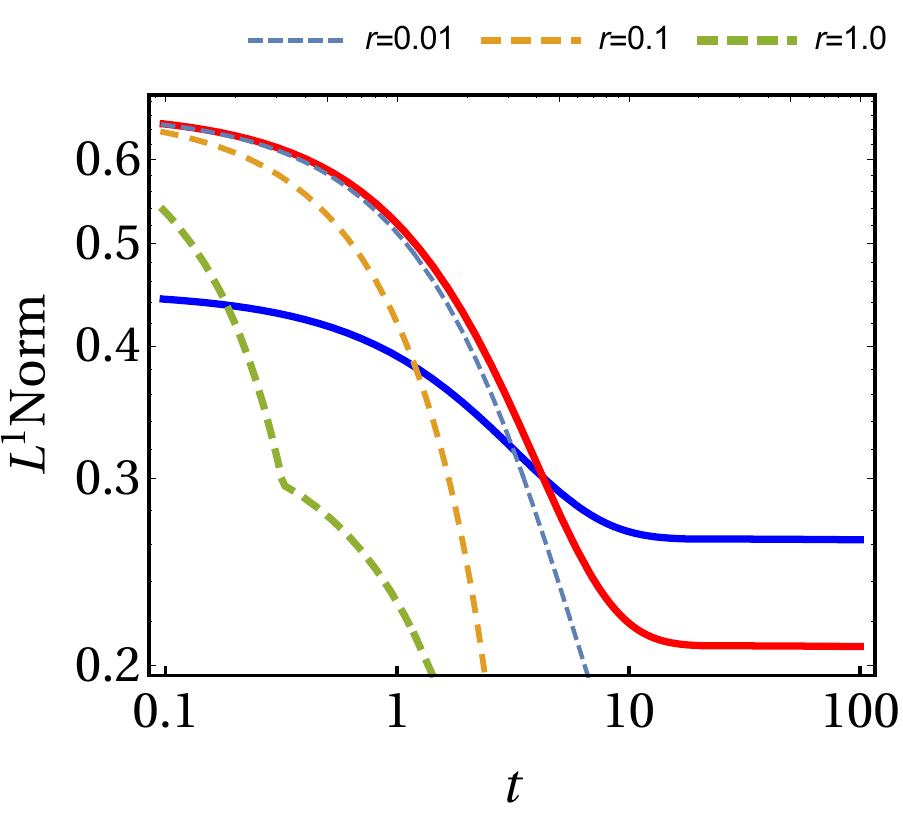}~~
    \includegraphics[width = 6cm]{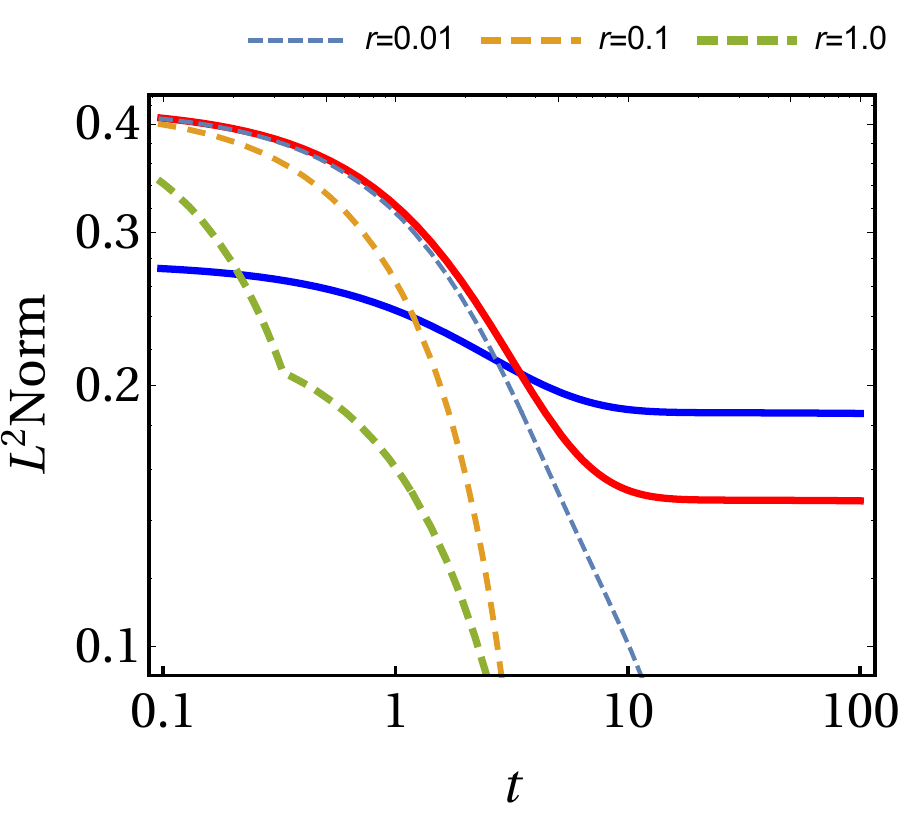}~~
    \includegraphics[width = 6cm]{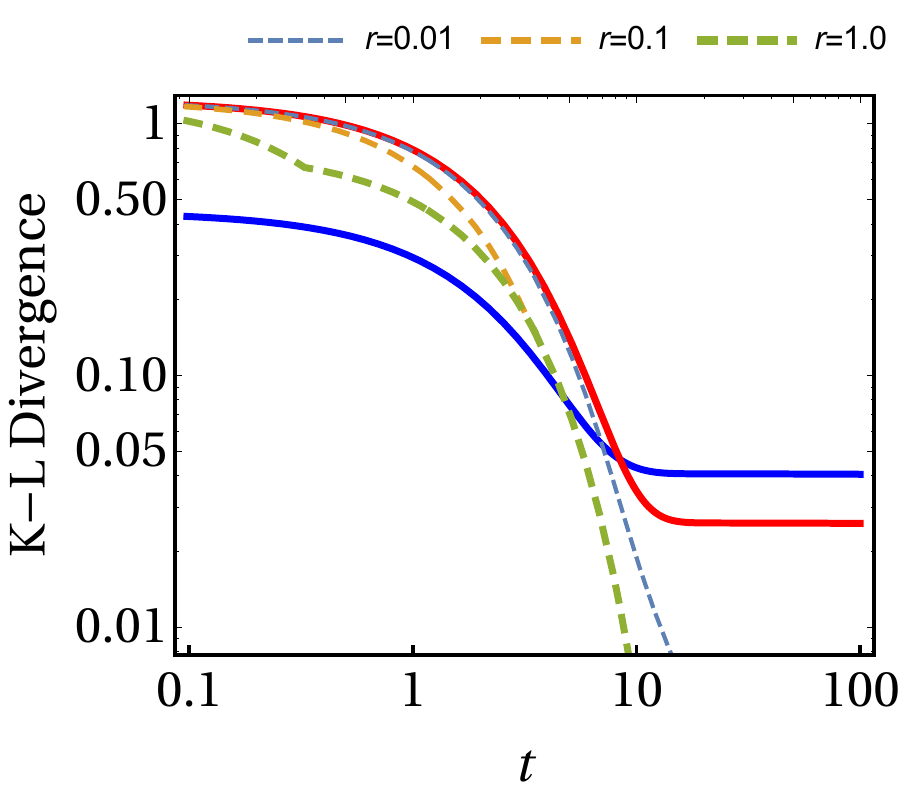}
    \caption{Comparison of different measures to observe the Mpemba effect (in log-log scale). Three different measures are used. First column: $L^1$norm. Second column: $L^2$norm, third column: KL divergence. In all plots, red and blue solid curves are, respectively, for distance of the time-dependent distribution, initialized from the Gibbs distribution at hot and cold temperature, from the equilibrium bath distribution (see main text). Dashed curves are the corresponding distances in the presence of stochastic resetting. Top panel: Resetting is inducing the Mpemba effect. Bottom Panel: Resetting reduces the time for onset of Mpemba effect. The parameters for all plots are $E_1=0.0, ~E_2=0.1,~E_3=0.7,~B_{1,2}=1.5,~B_{1,3}=0.8,~B_{2,3}=1.2,~T_{\rm b}=0.1$. For top panel: $T_{\rm H}=0.6, ~T_{\rm C}=0.15$ and for bottom panel: $T_{\rm H}=1.3, ~T_{\rm C}=0.42$. }
    \label{fig:S1}
\end{figure}

\section{Different measures do not qualitatively change the behaviors}
\label{kl-div-sec}
Here, we show that different measures do not qualitatively change the features discussed in the main text. In particular, we compare the $L^1$-norm, the $L^2$-norm (used throughout the manuscript), and the Kullback-Leibler divergence, in two different situations. In Fig.~\ref{fig:S1}(top panel), we investigate the case in which resetting mechanism induces the Markovian Mpemba effect, while in Fig.~\ref{fig:S1}(bottom panel), we report the results for case in which the latter is accelerated by introducing the stochastic reset.

\section{{\color{black}Explicit calculation for Mpemba effect} for two-state model in the absence of resetting}
\label{twolvsec}
Here we consider the two-state model and compute the coefficients $a_2^{\rm (I)}$ {\color{black}(defined below Eq.~\eqref{IC})} exactly and see if the condition \eqref{cond-1} is satisfied. For the two-state model, the solution of the master equation \eqref{ME} is
\begin{align}
    \vec p(t) = \vec p^{\rm eq} + a_2^{\rm (I)} \vec v_2 e^{\lambda_2 t},\label{twolev}
\end{align}
where the eigenvalue $\lambda_2 \equiv -(w^{\rm eq}_{12}+w^{\rm eq}_{21})<0$ and the corresponding right eigenvector $\vec v_2 \equiv \dfrac{1}{\sqrt{2}}\begin{pmatrix}1\\-1\end{pmatrix}$. Therefore, 
\begin{align}
    a_2^{\rm (I)}=\sqrt{2} [p_1^{\rm (I)}-p_{1}^{\rm eq}].\label{a2H}
\end{align}
is obtained by substituting the initial condition $\vec p(0)=\begin{pmatrix}p_1^{\rm (I)}\\p_2^{\rm (I)}\end{pmatrix}$ and $\vec p^{\rm (\eq)}=\begin{pmatrix}p_1^{\rm eq}\\p_2^{\rm eq}\end{pmatrix}$ in Eq.~\eqref{twolev}. 

Therefore, we have
\begin{align}
    \dfrac{a_2^{\rm (H)}}{a_2^{\rm (C)}}=\dfrac{1-e^{-(\beta_{\rm b}-\beta_{\rm H}) \Delta E}}{1-e^{-(\beta_{\rm b}-\beta_{\rm C}) \Delta E}} \times \dfrac{1+e^{\beta_{\rm H} \Delta E}}{1+e^{\beta_{\rm C} \Delta E}},
\end{align}
for the difference of energy $\Delta E \equiv E_2 -E_1$ between states 2 and 1. Defining $x \equiv e^{\beta_{\rm b} \Delta E}$, $y \equiv e^{\beta_{\rm H} \Delta E}$, and $z \equiv e^{\beta_{\rm C} \Delta E}$, where we have $y<z<x$, the above equation becomes
\begin{align}
    \dfrac{a_2^{\rm (H)}}{a_2^{\rm (C)}}=\dfrac{(x-y)(1+z)}{(x-z)(1+y)} > 1,\label{phs}
\end{align}
since $x-y>x-z>0$ and $1+z>1+y$. This eventually dictates that $|a_2^{\rm (H)}|> |a_2^{\rm (C)}|$.

Therefore, in any two-state system, the Markovian Mpemba effect does not exist. Therefore, to observe this effect, we should have at least three states.

\section{Strong Mpemba line in the presence of resetting}
\label{SM-sec-app}
In section~\ref{MME} for resetting rate $r=0$, we have seen that when the absolute value of the second coefficient of the expansion for the system initialized at hot temperature is lower than that of the same system initialized at cold temperature \eqref{cond-1}, we achieve the Markovian Mpemba effect. In this case, the system follows a trajectory in the probability space, and the system will eventually hit the bath temperature distribution $\vec p^{\rm eq}$ starting from either a hot $T_{\rm H}$ or a cold temperature $T_{\rm C}$ Gibbs-Boltzmann distribution. Conversely, in the presence of resetting, we stochastically reset the system to a particular state for a certain interval of time from the beginning, and the system can explore a larger portion of the probability space. 

The \textit{strong Mpemba space} (a line for three-state systems) is defined as the set of points in the probability space such that $|a_2^{\rm (H)}| = 0$, so that the hot system will relax to the bath temperature exponentially faster than the cold system. Using stochastic reset, we can drive the system, towards this space, and after that the system hits it, we switch off the resetting, and the system follows a reset-free dynamics.

In the following, we consider a system initialized at the equilibrium distribution associated to the hot temperature, $\vec{p}({T_{\rm H}})$. Hence, in Eq.~\eqref{anRt}, $a_n \equiv a_n^{\rm (H)}$, which are the coefficients associated with the starting distribution. The condition to hit the Strong Mpemba space at $t=t_{\rm SM}$ is
\begin{equation}
    \left( a_2^{(\rm H)} - \frac{r d_2}{r - \lambda_2} \right) e^{-r t_{\rm SM}} + \frac{r d_2}{r - \lambda_2} e^{-\lambda_2 t_{\rm SM}}  = 0.
    \label{SMLeq}
\end{equation}
Thus, for any fixed resetting rate $r$, and resetting state (encoded in $\vec{\Delta}$), we obtain the time at which the system hits the Strong Mpemba space, $t_{\rm SM}$:
\begin{align}
    t_{\rm SM}(r)=\dfrac{1}{|\lambda_2|+r}\ln \bigg[1-\dfrac{a_2^{(\rm H)}(|\lambda_2|+r)}{rd_2}\bigg].\label{time-sm-3}
\end{align}
Clearly, when $r\to \infty$, the time to reach the strong Mpemba line becomes zero. 

Therefore, in order to have $t^{\rm SM}(r)\geq0$, i.e., the system can hit the Strong Mpemba space in finite time for a given resetting rate and a given resetting state, we should impose the condition
\begin{align}
\dfrac{a_2^{(\rm H)}}{d_2}\leq 0.\label{cond-2}
\end{align}
In other words, we can actually find some states to reset the system, so that it will hit the strong Mpemba space in finite time $t_{\rm SM}(r)$. Interestingly, given a set of states for which strong Mpemba line exists, we can also find the optimal state depending on time to reach that space. This can be done as follow. Let $\{a_2^{(\rm H)}/d_2\}$ be the set of all ratios where each negative number corresponds to one particular resetting state. Evidently, the largest among them will give us the smaller time to reach strong Mpemba line for a given $r$ [see Eq.~\eqref{time-sm-3}].

\section{{\color{black}Explicit calculation for Mpemba effect} for two state model under resetting}
\label{two-state-app}
Here we consider a two-state system under resetting and we find a state which is the optimal one for resetting to reach the strong Mpemba line. Let us first expand the resetting vector $\vec\Delta$ in terms of eigenvectors of the $\hat{W}^{\rm eq}$ matrix 
\begin{align}
    \vec{\Delta} = \vec p^{\rm eq} +d_2 \vec v_2, 
\end{align}
where $\vec v_2 \equiv \dfrac{1}{\sqrt{2}}\begin{pmatrix}1\\-1\end{pmatrix}$.
Solving for $d_2$, we find
\begin{align}
    d_2(\ell) = \sqrt{2} [p_2^{\rm eq}\delta_{\ell,1}-p_1^{\rm eq}\delta_{\ell,2}],\label{d2l}
\end{align}
where $\ell = 1$ or $2$ is the resetting state, and $p_i^{\rm eq}$ is the equilibrium distribution of state $i$ at temperature $T_{\rm b}$. Clearly, $d_2(2)<0$, while $d_2(1)>0$.

Now, for the strong Mpemba space to exist, the resetting state should satisfy the condition \eqref{cond-2}. To identify the sign of the ratio of the left hand side of \eqref{cond-2}, we first look at the numerator, i.e.,  $a_2^{(\rm H)}$, obtained by substituting I by H in Eq.~\eqref{a2H}.
Thus, we have
\begin{align}
a_2^{(\rm H)} &\equiv \sqrt{2}[p_1(T_{\rm H})-p_{1}^{\rm eq}]\\ 
 &=\dfrac{e^{-\beta_{\rm H} \Delta E}\bigg[e^{-(\beta_{\rm b}-\beta_{\rm H}) \Delta E}-1]}{(1+e^{-\beta_{\rm H} \Delta E})(1+e^{-\beta_{\rm b} \Delta E})} < 0.
\end{align}

Thus, the condition \eqref{cond-2} is satisfied [by using Eq.~\eqref{d2l}].
\begin{align}
    \dfrac{a_2^{(\rm H)}}{d_2(\ell)}< 0
\end{align}
only if we reset to state 1 (see $d_2(\ell)$). 

In summary, from Sec.~\ref{twolvsec} we find that that there is no Markovian Mpemba effect for the two-level system. Notwithstanding, using resetting mechanism, whereby resetting the system to the lowest energy state for a time $t_{\rm SM}(r)$, we hit the strong Mpemba space, hence inducing the Markovian Mpemba effect in any two state system.

\section{{\color{black}Explicit calculation for Mpemba effect} for two-state model under resetting {\color{black}to} a mixed state}
\label{mix-state}
In section \ref{twolvsec}, we show that the Mpemba effect does not occur for the two-state model. We also derive the condition to induce the Markovian Mpemba effect with resetting to one single state {\color{black}in the previous section}. Here, we consider the slightly more complicated case in which we can reset to an ensemble of resetting locations, encoded in $\vec \Delta$ (see main text):
\begin{align}
    \vec\Delta = \begin{pmatrix}
    \mu\\
    1-\mu
    \end{pmatrix},
\end{align}
where $\mu \in[0,1]$ and $1-\mu$, respectively, are the probabilities to reset the system to state 1 and 2. Here $\mu=1$ ($\mu$=0) means that we reset to only state 1 (state 2). This resetting vector $\vec \Delta$ can be expanded in the eigenbasis of $\hat W^{\rm eq}$, i.e.,
\begin{align}
    \vec \Delta =\vec p^{\rm \eq} + d_2 \vec v_2,
\end{align}
where the eigenvector $\vec v_2$ is given in Sec. \ref{twolvsec}, and $d_2$ is a coefficient yet to be determined.  Substituting the value of $\vec v_2$ and $\vec p^{\rm eq}$, the coefficient $d_2$ reads
\begin{align}
    d_2 = \sqrt{2}\bigg(\mu - \dfrac{1}{1+e^{-\beta \Delta E}}\bigg),\label{d2eqn}
\end{align}
where $\Delta E \equiv E_2-E_1>0$ is the energy difference between state 2 and 1. 

\begin{figure}[t]
    \centering
    \includegraphics[width = 8 cm]{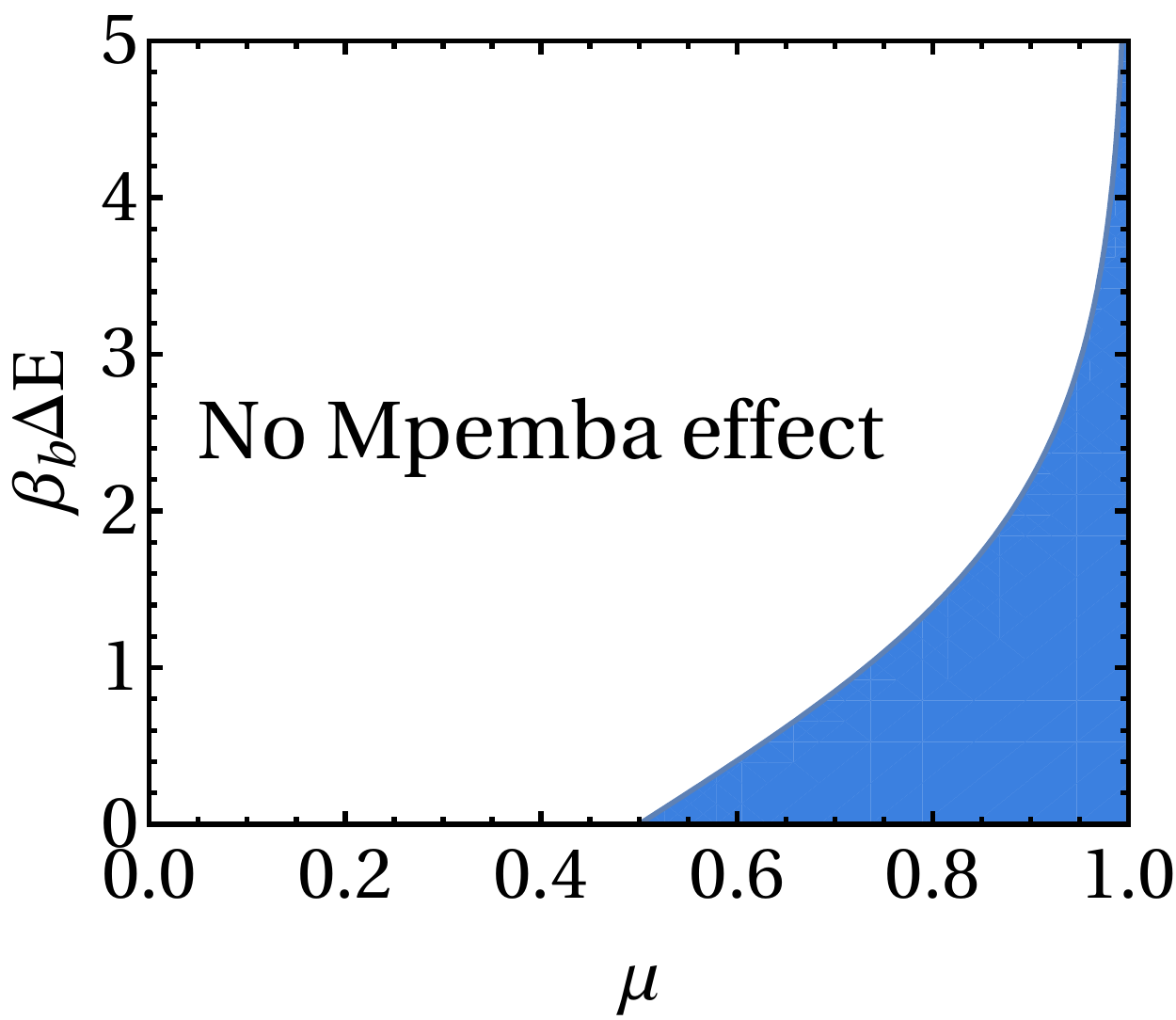}
    \caption{Mpemba effect for two-state model under stochastic reset. Phase diagram in the $(\mu,\beta_{\rm b} \Delta E)$ space is shown using the condition given in Eq.~\eqref{cond-3}. The shaded area corresponds to region where the Mpemba effect can be observed under stochastic resetting.  }
    \label{2-state}
\end{figure}

The condition for the onset of the Mpemba effect is given in Eq.~\eqref{cond-2}. To identify the sign of the ratio of the left-hand side of \eqref{cond-2}, we first look at the numerator, i.e.,  $a_2^{(\rm H)}$ obtained by substituting ${\rm I}$ by ${\rm H}$ in Eq.~\eqref{a2H}. Thus, we have
\begin{align}
a_2^{(\rm H)} &\equiv \sqrt{2}[p_1^{(T_{\rm H})}-p_{1}^{\rm eq}]\\ 
 &=\dfrac{e^{-\beta_{\rm H} \Delta E}\bigg[e^{-(\beta_{\rm b}-\beta_{\rm H}) \Delta E}-1]}{(1+e^{-\beta_{\rm H} \Delta E})(1+e^{-\beta_{\rm b} \Delta E})} < 0
\end{align}
since $\beta_{\rm b}>\beta_{\rm H}$ and $\Delta E>0.$
Therefore, the sign of the ratio \eqref{cond-2} will be negative if $d_2>0$, i.e., 
for (see Eq.~\eqref{d2eqn}) \begin{align}
    \mu > \dfrac{1}{1+\exp(-\beta_{\rm b} \Delta E)},
\end{align}
which can be rewritten as 
\begin{align}
    \beta_{\rm b} \Delta E < \ln\dfrac{\mu}{1-\mu},\label{cond-3}
\end{align}
where the left-hand side is the dimensionless energy difference between state 2 and 1.

In Fig.~\ref{2-state}, we present the phase diagram in the $(\mu,\beta_{\rm b} \Delta E$) space indicating when the condition~\eqref{cond-3} for the Mpemba effect occurs in the presence of stochastic resetting. Clearly, when $\mu \to 1$, we have that the effect can be induced independently of $\Delta E$, however the phase diagram is much more rich in this more complicated scenario.

\section{Pareto front for a three-state fully connected system}
\label{3-state-PF}
Consider a three-state fully connected system, specified by the following model parameters: $E_1 = 0$, $E_2 = 0.16$, $E_3 = 0.6$, $B_{1,2} = 0.7$, $B_{1,3} = 0.1$, $B_{2,3} = 1.13$. We set $T_H = 1.94$, and $T_C = 0.95$, with $T_b = 0.1$. We can draw a Pareto front for this system as discussed in the main text. Here, the front has no jump, since the optimal resetting state is always equal to $i_r^{\rm opt} = 1$, hence we did not include a smoothing parameter $\mu$ (as for a four-state system).

\begin{figure}[t]
    \centering
    \includegraphics[width = 0.75\columnwidth]{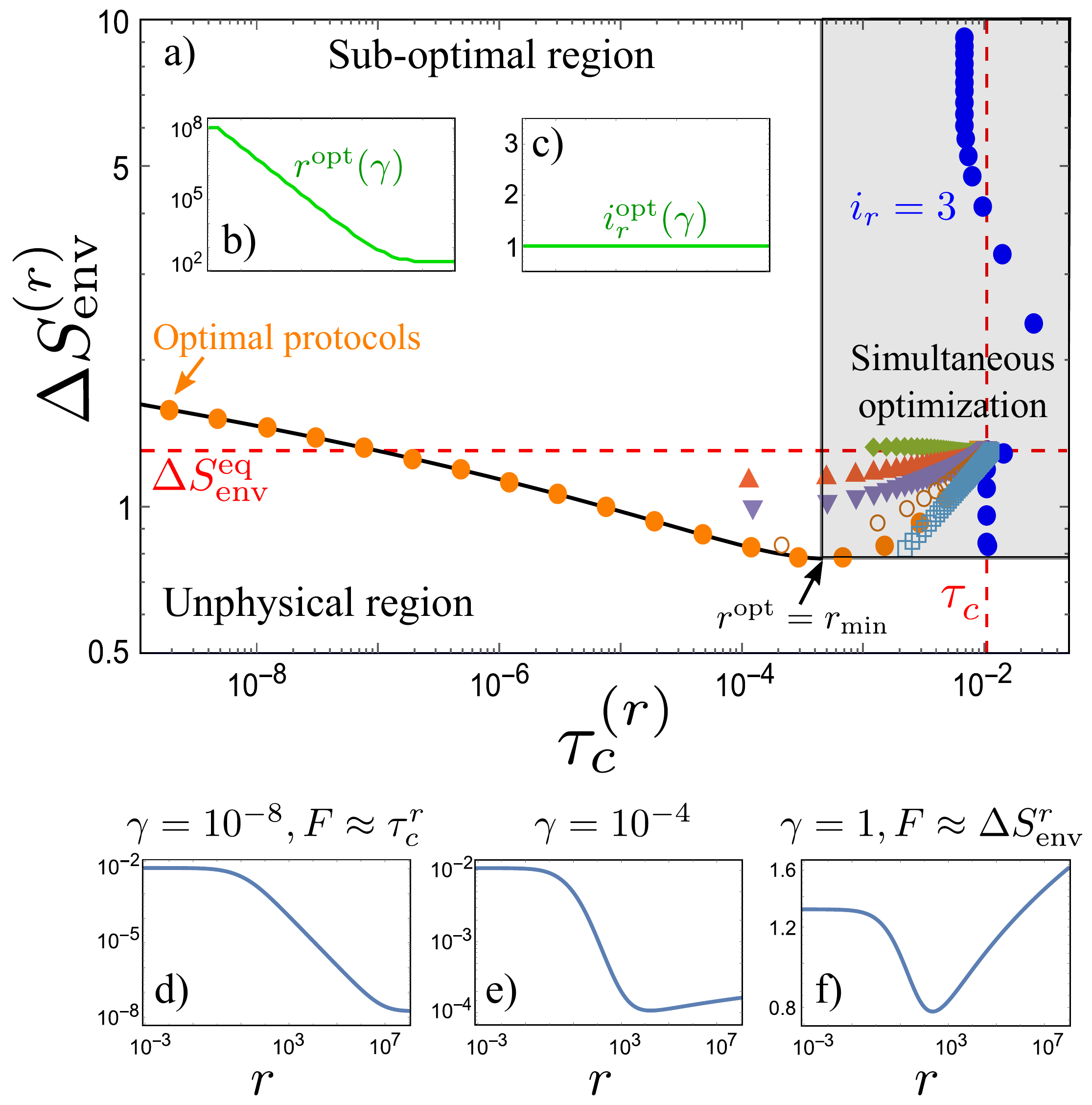}
    \caption{a) Pareto front for a three-state fully connected system (solid black line). Orange points are obtained setting $i_r = i_r^{\rm opt} = 1$. All the points in the gray-shaded area can be simultaneously optimized in the ($\Delta S_{\rm env}, \tau_{\rm c}^{(r)})$ space. Circles, squares, and triangles represent sub-optimal protocols obtained setting $t_r < \tau_{\rm c}^{(r)}$. b) Optimal resetting rate. c) Optimal resetting state, which coincides with the lowest energy state. d-f) Functional $F$ (see main text) for different values of $\gamma$, evaluated at $i_r = 1$ for simplicity. In particular, in f) $F \approx \Delta S_{\rm env}$, showing the presence of a minimum at $r = r_{\rm min}.$}
    \label{fig:pareto}
\end{figure}

In Fig.~\ref{fig:pareto}a, the black solid line represents the Pareto front. The orange dots are optimal protocols obtained for $i_r = i_r^{\rm opt}(\gamma) = 1$. The blue points have been obtained setting $i_r = 3$. Squares, triangles and circles indicates sub-optimal protocols obtained by stopping the resetting at a time $t_r < \tau_{\rm c}^{(r)}$, in contrast to the optimal protocol reported in the main text. Finally, the gray-shaded area indicates the points susceptible to simultaneous optimization of $\Delta S_{\rm env}$ and $\tau_{\rm c}^{(r)}$. In other words, all the protocols lying in this region, when optimized, collapse to one single point on the Pareto front, which is the one characterized by $r^{\rm opt} = r_{\rm min}$.

\section{Parameters for figures in the main text}
\label{parameters}
Here, we report the set of parameters to obtain the figures shown in the main text:
\begin{itemize}
\item[Fig.~1)] We set $E_1 = 0$, $E_2 = 0.1$, $E_3 = 0.6$, $B_{1,2} = 0.8$, $B_{1,3} = 1.2$, and $B_{2,3} = 1.13$. The fixed temperatures are $T_{\rm H} = 4$, $T_{\rm C} = 0.8$, and $T_{\rm b} = 0.1$. The resetting rate is $r = 10^2$.
\item[Fig.~2)] We set $E_1 = 0$, $E_2 = 0.08$, $E_3 = 0.6$, $E_4 = 1.1$, $B_{1,2} = 0.8$, $B_{1,3} = 1.5$, $B_{1,4} = 0.3$, $B_{2,3} = 0.5$, $B_{2,4} = 1.9$, and $B_{3,4} = 1.9$. The fixed temperatures are $T_{\rm H} = 1.7$, $T_{\rm C} = 0.8$, and $T_{\rm b} = 0.1$.
\end{itemize}

\end{document}